\PassOptionsToPackage{unicode}{hyperref}
\PassOptionsToPackage{hyphens}{url}
\PassOptionsToPackage{dvipsnames,svgnames,x11names}{xcolor}
\documentclass[
  12pt,
]{article}
\usepackage{amsmath,amssymb}
\usepackage{iftex}
\ifPDFTeX
  \usepackage[T1]{fontenc}
  \usepackage[utf8]{inputenc}
  \usepackage{textcomp} 
\else 
  \usepackage{unicode-math} 
  \defaultfontfeatures{Scale=MatchLowercase}
  \defaultfontfeatures[\rmfamily]{Ligatures=TeX,Scale=1}
\fi
\usepackage{lmodern}
\ifPDFTeX\else
\fi
\IfFileExists{upquote.sty}{\usepackage{upquote}}{}
\IfFileExists{microtype.sty}{
  \usepackage[]{microtype}
  \UseMicrotypeSet[protrusion]{basicmath} 
}{}
\usepackage{xcolor}
\usepackage[margin=1in]{geometry}
\usepackage{graphicx}
\makeatletter
\def\maxwidth{\ifdim\Gin@nat@width>\linewidth\linewidth\else\Gin@nat@width\fi}
\def\maxheight{\ifdim\Gin@nat@height>\textheight\textheight\else\Gin@nat@height\fi}
\makeatother
\setkeys{Gin}{width=\maxwidth,height=\maxheight,keepaspectratio}
\makeatletter
\def\fps@figure{htbp}
\makeatother
\usepackage{setspace}  \usepackage{float} \usepackage{caption}  
\ifLuaTeX
  \usepackage{selnolig}  
\fi
\usepackage[]{natbib}
\IfFileExists{bookmark.sty}{\usepackage{bookmark}}{\usepackage{hyperref}}
\IfFileExists{xurl.sty}{\usepackage{xurl}}{} 
\hypersetup{
  pdfauthor={Quang Nguyen; Gregory J. Matthews},
  colorlinks=true,
  linkcolor={cyan},
  filecolor={Maroon},
  citecolor={cyan},
  urlcolor={cyan},
  pdfcreator={LaTeX via pandoc}}

\title{Filling the Gaps: A Multiple Imputation Approach to Estimating
Aging Curves in Baseball}
\author{Quang
Nguyen\footnote{Department of Statistics \& Data Science, Carnegie Mellon University} \and Gregory
J.
Matthews\footnote{Department of Mathematics and Statistics, Loyola University Chicago}}
\date{March 11, 2024}

\begin{document}
\maketitle
\begin{abstract}
In sports, an aging curve depicts the relationship between average
performance and age in athletes' careers. This paper investigates the
aging curves for offensive players in Major League Baseball. We study
this problem in a missing data context and account for different types
of dropouts of baseball players during their careers. We employ a
multiple imputation framework for multilevel data to impute the player
performance associated with the missing seasons, and estimate the aging
curves based on the imputed datasets. We then evaluate the effects of
different dropout mechanisms on the aging curves through simulation,
before applying our method to analyze MLB player data from past seasons.
Results suggest an overestimation of the aging curves constructed
without considering the unobserved seasons, whereas estimates obtained
from multiple imputation address this shortcoming.\\
\strut 
\end{abstract}

\newpage

\hypertarget{sec:intro}{%
\section{Introduction}\label{sec:intro}}

The rise and fall of an athlete is a popular topic of discussion in the
sports media today. Questions regarding whether a player has reached
their peak, is past their prime, or is good enough to remain in their
respective professional league are often seen in different media outlets
such as news articles, television debate shows, and podcasts. The
average performance of players by age throughout their careers is
visually represented by an \emph{aging curve}. This graph typically
consists of a horizontal axis representing a time variable (usually age
or season) and a vertical axis showing a performance metric at each time
point in a player's career.

One significant challenge associated with the study of aging curves in
sports is \emph{survival bias}, as pointed out by
\citet{Lichtman2009baseball}, \citet{Turtoro2019flexible},
\citet{Judge2020approach}, and \citet{Schuckers2023observed}. In
particular, the aging effects are not often determined from a full
population of athletes (i.e., all players who have ever played) in a
given league. That is, only players that are good enough to remain are
observed; whereas those who might be involved, but do not actually
participate or are not talented enough to compete, are being completely
disregarded. This very likely results in an overestimation of the aging
curves.

As such, player survivorship and dropout can be viewed as a missing data
problem. There are several distinct cases of player absence from
professional sport at different points in their careers. Early on, teams
may elect to assign their young prospects to their minor/development
league affiliates for several years of nurture. Many of those players
would end up receiving a call-up to join the senior squad, when the team
believes they are ready. During the middle of a player's career, a
nonappearance could occur due to various reasons. Injury is unavoidable
in sports, and this could cost a player at least one year of their
playing time. Personal reasons such as contract situation and more
recently, concerns regarding a global pandemic, could also lead to
athletes sitting out a season. Later on, a player, by either their
choice or their team's choice, might head for retirement because they
cannot perform at a level like they used to, leading to unobserved
seasons that could have been played.

The primary aim of this paper is to apply missing data techniques to the
estimation of aging curves. In doing so, we focus on baseball and pose
the following research question: \textit{What would the aging curve look
like if all players competed in every season within a fixed range of
age?} In other words,
\textit{what would have happened if a player who was
forced to retire from their league at a certain age had played a full
career?} The manuscript continues with a review of existing literature
on aging curves in baseball and other sports in Section \ref{sec:lit}.
Next, we describe our data and methods used to perform our analyses in
Section \ref{sec:meth}. After that, our approach is implemented through
simulation and analyses of real baseball data in Sections \ref{sec:sim}
and \ref{sec:app}. Finally, in Section \ref{sec:discuss}, we conclude
with a discussion of the results, limitations, and directions for future
work.

\hypertarget{sec:lit}{%
\section{Literature Review}\label{sec:lit}}

To date, we find a considerable amount of previous work related to aging
curves and career trajectory of athletes. This body of work consists of
several key themes, a wide array of statistical methods, and
applications in many sports besides baseball such as basketball, hockey,
and track and field, to name a few.

A typical notion in the baseball aging curves literature is the
assumption of a quadratic form for modeling the relationship between
performance and age. \citet{Morris1983parametric} looked at Ty Cobb's
batting average trajectory using parametric empirical Bayes and used
shrinkage methods to obtain a parabolic curve for Cobb's career
performance. \citet{Albert1992bayesian} proposed a quadratic random
effects log-linear model for smoothing a batter's home run rates
throughout their career. \citet{Berry1999bridging} implemented a
nonparametric method to estimate the age effect on performance in
baseball, hockey, and golf. However, \citet{Albert1999comment} weighed
in on this nonparametric approach and questioned the assumptions that
the peak age and periods of growth and decline are the same for all
players. \citet{Albert1999comment} ultimately preferred a second-degree
polynomial function for estimating age effect in baseball, which is a
parametric model. Continuing his series of work on aging trajectories,
\citet{Albert2002smoothing} proposed a Bayesian exchangeable model for
baseball hitting performance. This approach combined quadratic
regression estimates and assumes similar careers for players born in the
same decade. \citet{Fair2008estimated} and \citet{Bradbury2009peak} both
used a fixed-effects regression to examine age effects in the MLB, also
assuming a quadratic aging curve form. A major drawback of
\citet{Bradbury2009peak}'s study is that the analysis only considered
players with longer baseball careers.

In addition to baseball, studies on aging curves have also been
conducted for other sports. Early on, \citet{Moore1975study} looked at
the association between age and running speed in track and field and
produced aging curves for different running distances using an
exponential model. \citet{Fair1994fast} and \citet{Fair2007estimated}
studied the age effects in track and field, swimming, chess, and
running, in addition to their latter work in baseball, as mentioned
earlier. In triathlon, \citet{Villaroel2011elite} assumed a quadratic
relationship between performance and age, as many have previously
considered. As for basketball, \citet{Page2013effect} used a Gaussian
process regression in a hierarchical Bayesian framework to model age
effect in the NBA. Additionally, \citet{Lailvaux2014trait} used NBA and
WNBA data to investigate and test for potential sex differences in the
aging of comparable performance indicators. \citet{Vaci2019large}
applied Bayesian cognitive latent variable modeling to explore aging and
career performance in the NBA, accounting for player position and
activity level. In tennis, \citet{Kovalchik2014older} studied age and
performance trends in men's tennis using change point analysis.

Another convention in the aging curve modeling literature is the
assumption of discrete observations. Specifically, most researchers use
regression modeling and consider a data measurement for each season
played throughout a player's career. In contrast to previous approaches,
\citet{Wakim2014functional} took a different route and considered
functional data analysis as the primary tool for modeling MLB and NBA
aging curves. This is a continuous framework which treats the entire
career performance of an athlete as a smooth function. In a similar
functional data setting, \citet{Leroy2018Functional} investigated the
performance progression curves in swimming.

A subset of the literature on aging and performance in sports
specifically studies the question: At what age do athletes peak?
\citet{Schulz1988peak} looked at the age of peak performance for track
and field, swimming, baseball, tennis, and golf. A follow-up study to
this work was done by \citet{Schulz1994relationship}, where the authors
focused on baseball and found that the average peak age for baseball
players is between 27 and 30, considering several performance measures.
Later findings on baseball peak age also showed consistency with the
results in \citet{Schulz1994relationship}. \citet{Fair2008estimated}
determined the peak-performance age in baseball to be 28, whereas
\citet{Bradbury2009peak} determined that baseball hitters and pitchers
reach the top of their careers at about 29 years old. In soccer,
\citet{Dendir2016soccer} determined that the peak age for footballers in
the top leagues falls within the range of 25 to 27.

The idea of player survivorship is only mentioned in a small number of
articles. To our knowledge, not many researchers have incorporated
missing data methods into the estimation of aging curves to account for
missing but observable athletes. \citet{Schulz1994relationship} and
\citet{Schell2005baseball} noted the selection bias problem with
estimating performance averages by age in baseball, as better players
tend to have longer career longevity. \citet{Schall2000career} predicted
survival probabilities of baseball players using a logit model, and
examined the link between first-year performance and career length.
\citet{Lichtman2009baseball} studied different aging curves for
different eras and groups of players after correcting for survival bias,
and showed that survival bias results in an overestimation of the age
effects. In other words, the average performance-level at any given age
is overestimated when survival bias is not considered.
\citet{Judge2020approach} also examined survival bias and found that the
slope between age and performance is underestimated when using only
surviving players in the estimation. We note that
\citet{Judge2020approach} studied how the slope of the aging curve is
affected by survival bias, whereas \citet{Lichtman2009baseball}
investigated the bias in the average performance level at each specific
age (i.e.~the height of the curve) which is similar to our goal in this
paper. In analyzing NHL player aging, \citet{Brander2014estimating}
applied their quadratic and cubic fixed-effects regression models to
predict performance for unobserved players in the data.

Recently, researchers have noticed the benefits of accounting for
missing data in modeling performance in sports.
\citet{Stival2023missing} used a latent class matrix-variate state-space
framework to analyze runners' careers, and found that missing data
patterns greatly contribute to the prediction of performance. Perhaps
the most closely related approach to our work is that by
\citet{Schuckers2023observed}, which considered different regression and
imputation frameworks for estimating the aging curves in the National
Hockey League (NHL). First, they investigated different regression
approaches including spline, quadratic, quantile, and a delta plus
method, which is an extension to the delta method previously studied by
\citet{Lichtman2009baseball}, \citet{Turtoro2019flexible}, and
\citet{Judge2020delta}. This paper also proposed an imputation approach
for aging curve estimation, and ultimately concluded that the estimation
becomes stronger when accounting for unobserved data, which addresses a
major shortcoming in the estimation of aging curves.
\citet{Safvenberg2022age} modified \citet{Schuckers2023observed}'s
imputation algorithm to study aging trajectory in Swedish football.

However, it appears that the aging curves in the aforementioned papers
are constructed without taking into account the variability as a result
of imputing missing data. This could be improved by applying multiple
imputation rather than considering only one imputed dataset. As pointed
out by \citet{Gelman2006data} (Chapter 25), conducting only a single
imputation essentially assumes that the filled-in values correctly
estimate the true values of the missing observations. Yet, there is
uncertainty associated with the missingness, and multiple imputation can
incorporate the missing data uncertainty and provide estimates for the
different sources of variability.

\hypertarget{sec:meth}{%
\section{Methods}\label{sec:meth}}

\hypertarget{sec:data}{%
\subsection{Data Collection}\label{sec:data}}

In the forthcoming analyses, we rely on one primary source of publicly
available baseball data: the Lahman baseball database
\citep{Lahman2021baseball}. Created and maintained by Sean Lahman, this
database contains pitching, hitting, and fielding information for Major
League Baseball players and teams dating back to 1871. The data are
available in many different formats, and the \texttt{Lahman} package in
\texttt{R} \citep{Friendly2022Lahman, R2023language} is used for our
investigation.

Due to our specific purpose of examining the aging curves for baseball
offensive players, we consider the following datasets from the
\texttt{Lahman} library: \texttt{Batting}, which provides
season-by-season batting statistics for baseball players; and
\texttt{People}, which contains the date of birth of each player,
allowing us to calculate their age for each season played. In each
table, an athlete is identified with their own \texttt{playerID}, hence
we use this attribute as a joining key to merge the two tables together.
A player's age for a season is determined as their age on June 30, and
we apply the formula suggested by \citet{marchi2018analyzing} for age
adjustment based on one's birth month.

Throughout this paper, we consider on-base plus slugging (OPS), which
combines a hitter's ability to reach base and power-hitting, as the
baseball offensive performance measure. We normalize the OPS for all
players and then apply an arcsine transformation to ensure a reasonable
range for the OPS values when conducting simulation and imputation. We
also assume a fixed length for a player's career, ranging from age 21 to
39. In terms of sample restriction, we observe all player-seasons with
at least 100 plate appearances, which means a season is determined as
missing if one's plate appearances is below that threshold.

\hypertarget{multiple-imputation}{%
\subsection{Multiple Imputation}\label{multiple-imputation}}

Multiple imputation \citep{Rubin1987multiple} is a popular statistical
procedure for addressing the presence of incomplete data. The goal of
this approach is to replace the missing data with plausible values to
create multiple completed datasets. These completed datasets can each
then be analyzed and results are combined across the imputed versions.
Multiple imputation consists of three steps. First, based on an
appropriate imputation model, \(m\) completed copies of the dataset are
created by filling in the missing values. Next, \(m\) analyses are
performed on each of the \(m\) completed datasets. Finally, the results
from each of the \(m\) datasets are pooled together using Rubin's rules
\citep{Little1987statistical}.

It is important to understand the reasons behind the missingness when
applying multiple imputation to handle incomplete data. Generally, there
are three types of missing data: missing completely at random (MCAR),
missing at random (MAR), and missing not at random (MNAR)
\citep{Rubin1976inference}. MCAR occurs when a missing observation is
statistically independent of both the observed and unobserved data. In
the case of MAR, the missingness is associated only with the observed
and not with the unobserved data. When data are MNAR, the probability of
missingness is related to both observed and unobserved data.

Among the tools for performing multiple imputation, multivariate
imputations by chained equation (MICE) \citep{Vanbuuren1999flexible} is
a flexible, robust, and widely used method. This algorithm imputes
missing data via an iterative series of conditional models. In each
iteration, each incomplete variable is filled in by a separate model of
all the remaining variables. The iterations continue until apparent
convergence is reached.

Here we implement the MICE framework in \texttt{R} via the popular
\texttt{mice} package \citep{vanBuuren2011mice}. Moreover, we focus on
multilevel multiple imputation, due to the hierarchical structure of our
data. Specifically, we consider multiple imputation by a two-level
normal linear mixed model with heterogeneous within-group variance
\citep{Kasim1998application}. In context, our data consist of baseball
seasons (ages) which are nested within the class variable, player; and
the season-by-season performance is considered to be correlated for each
athlete. The described imputation model can be specified as the
\texttt{2l.norm} method available in the \texttt{mice} library.

\hypertarget{sec:sim}{%
\section{Simulation}\label{sec:sim}}

In this simulation, we demonstrate our aging curve estimation approach
with multiple imputation, and evaluate how different types of player
dropouts affect the curve. There are three steps to our simulation.
First, we fit a model for the performance-age relationship and use its
output to generate reasonable careers for baseball players. For
computational simplicity, we consider a mixed-effects model that assumes
the same aging curve for all players and allowing for a random shift
across players. Next, we generate missing data by dropping players from
the full dataset based on different criteria, and examine how the
missingness affects the original aging curve obtained from fully
observed data. Finally, we apply multiple imputation to obtain completed
datasets and assess how close the imputed aging curves are to the true
curve based on fully observed data.

\hypertarget{generating-player-careers}{%
\subsection{Generating Player Careers}\label{generating-player-careers}}

We fit a mixed-effects model using the player data described in Section
\ref{sec:data}. Our goal is to obtain the variance components of the
fitted model to simulate baseball player careers. The model of interest
is of the following form: \[
\displaylines{
Y_{pq} = (\beta_0 + b_{0p}) + \beta_1X_q + \beta_2X_q^2 + \epsilon_{pq} 
\cr
b_{0p} \sim N(0, \tau^2) 
\cr 
\epsilon_{pq} \sim N(0, \sigma^2).
}
\] In detail, this model relates the performance metric \(Y_{pq}\) (in
our case, transformed OPS) for player \(p\) at age (season) \(q\) to a
baseline level via the fixed effect \(\beta_0\). The only covariate
\(X\) in the model is age, which, similar to previous work mentioned in
Section \ref{sec:lit}, is assumed to have a quadratic relationship with
the response variable, transformed OPS. Another component is the
observational-level error \(\epsilon_{pq}\) with variance \(\sigma^2\)
for player \(p\) at age \(q\). We also introduce the random effects
\(b_{0p}\), which represents the deviation from the grand mean
\(\beta_0\) for player \(p\), allowing a fixed amount of shift to the
performance prediction for each player. In addition, to incorporate the
variability in production across the season \(q\), a random effect
parameter \(\tau^2\) is included. Our modeling approach is implemented
using the \texttt{lme4} package in \texttt{R} \citep{Bates2015lme4}. We
use the estimated variance components from the fitted model to simulate
1000 careers for baseball players from the ages of 21 to 39.

\hypertarget{sec:drop}{%
\subsection{Generating Missing Data}\label{sec:drop}}

After obtaining reasonable simulated careers for baseball players, we
create different types of dropouts and examine how they lead to
deviations from the fully observed aging curve. We consider the
following cases of player dropout from the league:

\begin{enumerate}
\def\labelenumi{(\arabic{enumi})}
\item
  Dropout players with 4-year OPS average below a certain threshold, say
  0.55.
\item
  Dropout players with OPS average from age 21 (start of career) to 25
  of less than 0.55.
\item
  25\% of the players randomly retire at age 30.
\item
  Players with OPS average below 0.55 between age 21 and 25 are not being promoted until age 25.
\end{enumerate}

For the first two scenarios, the missingness mechanism is MAR, since
players get removed due to low previously observed performance.
Specifically, for case (2), all player-seasons at the beginning at their
career between age 21 and 25 are observed, and low-performing players
during this playing period are dropped. Dropout case (3) falls under
MCAR, since athletes are selected completely at random to retire without
considering past or future offensive production. Case (4) illustrates
MNAR, since our consideration is that players who have low OPS at the
beginning do not make it into the major league until they turn 25. Here,
the dropped OPS values are what their OPS would have been if they
played, but because it would have been so low, they do not get promoted
until the age cutoff of 25.

\begin{figure}

{\centering \includegraphics{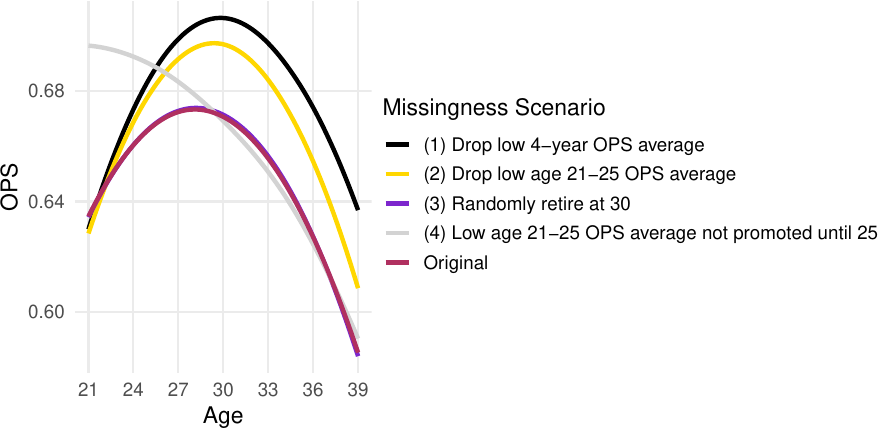} 

}

\caption{Comparison of the average aging curves constructed with the full simulated data (maroon) and different cases of dropouts (without imputation). The dropout mechanisms presented are (purple) randomly removing 25\% of the players at age 30; (black) dropping players with any 4-year OPS average below 0.55; (gold) dropping players with OPS average between age 21 and 25 of less than 0.55; and (gray) players with low OPS average between age 21 and 25 are not being promoted until age 25. Results shown here are for 1,000 player-careers generated from simulation.}\label{fig:drop-compare}
\end{figure}

Figure \ref{fig:drop-compare} displays the average OPS aging curve for
all baseball players obtained from the original data with no
missingness, along with the aging curves constructed based on data with
only the surviving players associated with the aforementioned dropout
mechanisms. For each dropout case, the corresponding curve is based on
the average OPS at each age aggregated across all players that remain
after dropout. These are smoothed curves obtained from loess model fits,
and we use mean absolute error (MAE) to evaluate the discrepancy between
the dropout and true aging curves.

It is clear that randomly removing players have minimal effect on the
aging curve, as the curve obtained from (3) and the original curve
essentially overlap (MAE \(= 7.42 \times 10^{-4}\)). On the other hand,
a positive shift from the fully observed curve occurs for the remaining
two cases of dropout based on OPS average (MAE \(= 0.031\) for (1) and
MAE \(=0.019\) for (2)). This means the aging curves with only the
surviving players are overestimated in the presence of missing data due
to past performance. More specifically, the estimated player performance
drops off faster as they age when considering missing data than when it
is estimated with only complete case analysis (i.e.~only considering
observed seasons). As for case (4), when low-performing players are not
being promoted until age 25 and only good players are observed before
age 25, we observe an overestimation at the beginning of player's
career. Aside from overestimating the aging curve, ignoring player
dropout also pushes the estimated performance peak to a later point (29
years old) in a player's career.

\hypertarget{sec:imp}{%
\subsection{Imputation}\label{sec:imp}}

Next, we implement the multiple imputation with a hierarchical structure
procedure described in Section \ref{sec:meth} to the cases of dropout
that shifts the aging effect on performance. We perform \(m=5\)
imputations with each made up of 30 iterations of the MICE algorithm,
and apply Rubin's rules for combining the imputation estimates. The
following results are illustrated for dropout mechanism (2), where
players with a low OPS average at the start of their careers (ages
21--25) are forced out of the league.

\begin{figure}

{\centering \includegraphics{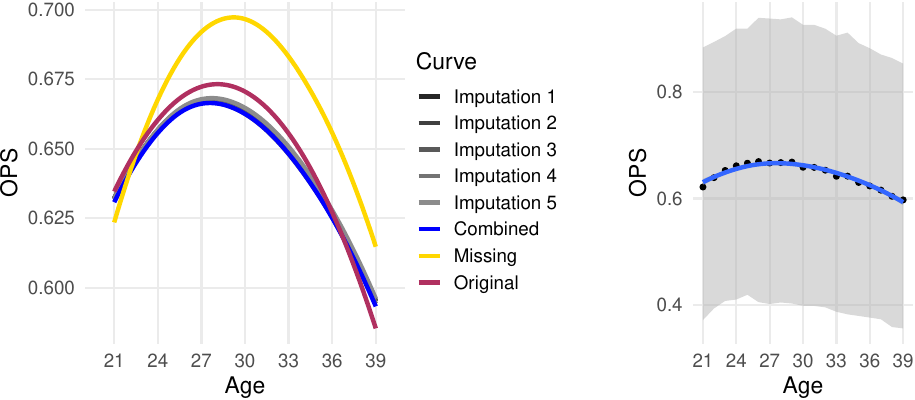} 

}

\caption{At left, comparison of the average OPS aging curves constructed with the fully observed data, only surviving players, and imputation. At right, combined imputed curve with 95\% confidence intervals obtained from Rubin's rules. Results shown here are for the dropout case of players having OPS average from age 21 to 25 below 0.55.}\label{fig:drop-imp}
\end{figure}

Figure \ref{fig:drop-imp} (left) shows smoothed fitting loess aging
curves for all 5 imputations and a combined version of them, in addition
to the curves constructed with fully observed and only surviving players
data. Similar to Figure \ref{fig:drop-compare}, these curves are
obtained by averaging the OPS for each age across all players with
non-missing OPS values. The 95\% confidence interval for the mean OPS at
each age point in the combined curve obtained from Rubin's rules is
further illustrated in Figure \ref{fig:drop-imp} (right). It appears
that the combined imputed curve follows the same shape as the true,
known curve. Moreover, imputation seems to capture the rate of change
for the beginning and end career periods quite well, whereas the middle
of career looks to be slightly underestimated. The resulting MAE of
\(0.0039\) confirms that there is little deviation of the combined curve
from the true one.

\begin{figure}

{\centering \includegraphics{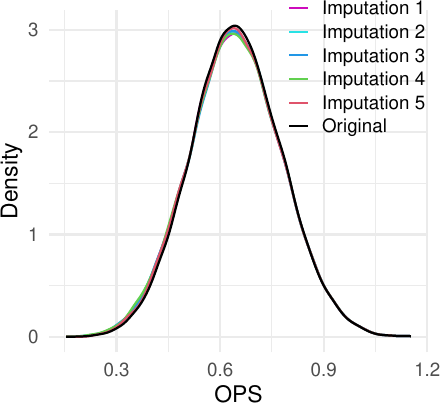} \includegraphics{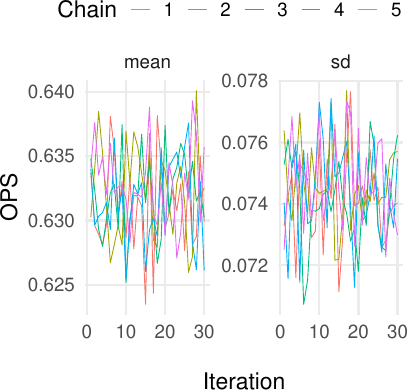} 

}

\caption{At left, kernel density estimates for the fully observed and imputed OPS values. At right, trace plots for the mean and standard deviation of the imputed OPS values against the iteration number for the imputed data. Results shown here are for the dropout case of players having OPS average from age 21 to 25 below 0.55.}\label{fig:diag}
\end{figure}

Additionally, we perform diagnostics to assess the plausibility of the
imputations, and also examine whether the MICE algorithm converges. We
first check for distributional discrepancy by comparing the
distributions of the fully observed and imputed data. Figure
\ref{fig:diag} (left) presents the density curves of the OPS values for
each imputed dataset and the fully simulated data. It is obvious that
the imputation distributions are well-matched with the observed data. To
confirm convergence of the MICE algorithm, we inspect trace plots for
the mean and standard deviation of the imputed OPS values. As shown in
Figure \ref{fig:diag} (right), convergence is apparent, since no
definite trend is revealed and the imputation chains are intermingled
with one another.

\hypertarget{sec:app}{%
\section{Application: MLB Data}\label{sec:app}}

Lastly, we apply the previously mentioned multilevel multiple imputation
model to estimate the average OPS aging curve for MLB players. For this
investigation, besides the data pre-processing tasks mentioned in
Section \ref{sec:data}, our sample is limited to all players who made
their major league debut no sooner than 1985, resulting in a total of
2323 players. To perform imputation, we pass in the same parameters to
our simulation study (\(m=5\) with 30 iterations for each imputation).

\begin{figure}

{\centering \includegraphics{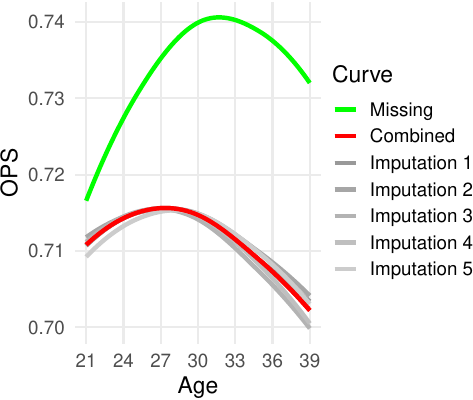} 

}

\caption{Comparison of the average OPS aging curves constructed with only observed players and imputation for MLB data. Note that the ``Missing'' curve is obtained when the missing data (unobserved player seasons) are ignored. Results shown here are for 2,323 players who made their MLB debut in or after 1985.}\label{fig:mlb-imp}
\end{figure}

Figure \ref{fig:mlb-imp} shows the OPS aging curves for MLB players
estimated with and without imputation. The plot illustrates a similar
result as our simulations as the combined curve based on imputations is
lower than the curve obtained when ignoring the missing data. The peak
age after imputation occurs much earlier (26 years old) compared to the
peak age when player dropout is not accounted for (30 years old). It is
clear that the aging effect is overestimated without considering the
unobserved player-seasons. In other words, the actual performance
declines with age more rapidly than estimates based on only the observed
data.

\hypertarget{sec:discuss}{%
\section{Discussion}\label{sec:discuss}}

The concept of survival bias is frequently seen in professional sports,
and our paper approaches the topic of aging curves and player dropout in
baseball as a missing data problem. We use multiple imputation with a
multilevel structure to improve estimates for the aging curves. Through
simulation, we highlight that ignoring the missing seasons leads to an
overestimation of the age effect on baseball offensive performance. With
imputation, we achieve an aging curve showing that players actually
decline faster as they get older than previously estimated.

There are notable limitations of our study which leave room for
improvement in future work. In our current imputation model, age is the
only predictor for estimating performance. It is possible to include
more covariates in the imputation algorithm and determine whether a
better aging curve estimate is achieved. In particular, we can factor in
other baseball offensive statistics (e.g., home run rate, strikeout
rate, WOBA, walk rate,\ldots) in building an imputation model for OPS.
We can also examine other performance metrics to see how age affects
different statistics.

Furthermore, the aging curve estimation problem can be investigated in a
completely different statistical setting. As noted in Section
\ref{sec:lit}, rather than considering discrete observations, another
way of studying aging curves is through a continuous approach, assuming
a smooth curve for career performance. As pointed out by
\citet{Wakim2014functional}, methods such as functional data analysis
(FDA) and principal components analysis through conditional expectation
(PACE) possess many modeling advantages, in regard to flexibility and
robustness. There exists a number of proposed multiple imputation
algorithms for functional data
\citep{He2011functional, Ciarleglio2021elucidating, Rao2021modern},
which all can be applied in future studies on aging curves in sports.

\hypertarget{acknowledgements}{%
\section*{Acknowledgements}\label{acknowledgements}}
\addcontentsline{toc}{section}{Acknowledgements}

We thank the organizers of the 2022 Carnegie Mellon Sports Analytics
Conference (CMSAC) for the opportunity to present this work and receive
feedback. We thank the anonymous reviewers of the Reproducible Research
Competition at CMSAC 2022 for the insightful comments and suggestions.
We thank Kathryne Piazza for her help in the early stages of this
project.

\hypertarget{supplementary-material}{%
\section*{Supplementary Material}\label{supplementary-material}}
\addcontentsline{toc}{section}{Supplementary Material}

\hypertarget{code-availability}{%
\subsection*{Code availability}\label{code-availability}}
\addcontentsline{toc}{subsection}{Code availability}

All code related to this paper is available at
\href{https://github.com/qntkhvn/aging}{\texttt{github.com/qntkhvn/aging}}.

\hypertarget{rubins-rules-for-pooling-parameter-estimates}{%
\subsection*{Rubin's rules for pooling parameter
estimates}\label{rubins-rules-for-pooling-parameter-estimates}}
\addcontentsline{toc}{subsection}{Rubin's rules for pooling parameter
estimates}

Let \(Q\) be a parameter of interest and \(\widehat Q_i\) where
\(i=1,2,\dots,m\) are estimates of \(Q\) obtained from \(m\) imputed
datasets, with sampling variance \(U\) estimated by \(\widehat U_i\).
Then the point estimate for \(Q\) is the average of the \(m\) estimates
\[
\overline Q = \frac{1}{m} \sum_{i=1}^m \widehat Q_i
\,.
\] The variance for \(\overline Q\) is defined as \[
T=\overline U + \left(1 + \frac{1}{m}\right)B
\,,
\] where \[
\overline U = \frac{1}{m} \sum_{i=1}^m \widehat U_i
\] and \[
B=\frac{1}{m-1} \sum_{i=1}^m (\widehat Q_i - \overline Q)^2
\] are the estimated within and between variances, respectively.

Inferences for \(Q\) are based on the approximation \[
\frac{Q - \overline Q}{\sqrt{T}} \sim t_\nu
\,,
\] where \(t_\nu\) is the Student's \(t\)-distribution with
\(\displaystyle \nu = (m-1)\left(1+\frac{1}{r}\right)^2\) degrees of
freedom, with
\(\displaystyle r=\left(1+\frac{1}{m}\right)\frac{B}{\overline U}\)
representing the relative increase in variance due to missing data.

Accordingly, a \(100(1-\alpha)\)\% Wald confidence interval for \(Q\) is
computed as \[
\overline Q \ \pm \ t_{\nu,1-\alpha/2}\sqrt{T}
\,,
\] where \(t_{\nu,1-\alpha/2}\) is the \(1-\alpha/2\) quantile of
\(t_\nu\).

\renewcommand\refname{References}
  \bibliography{references.bib}

\end{document}